\documentclass [a4paper,fleqn, 12pt]{article}
\usepackage{graphicx}
\usepackage[small]{subfigure,epsfig}

\usepackage {amsmath} \usepackage{amssymb} \usepackage{cite}

\newcommand{\cn}

\begin{document}


\title
{A Note on "M -- component nonlinear evolution equations: multiple soliton solutions"}

\author
{Naum N. Muraved}

\date{Institute of Computer Science, National Research Nuclear University
MEPHI, 31 Kashirskoe Shosse, 115409 Moscow, Russian Federation}

\maketitle

\begin{abstract}

We analyze the recent paper by Wazwaz [Wazwaz A.M., M -- component nonlinear evolution equations: multiple soliton solutions, Phys. Scr. 81 (2010) 055004]. We demonstrate that author did not consider in essence the M -- component nonlinear evolution equations but he reduced the M -- component equations to the well -- known Korteweg -- de Vries equation, to modified Korteweg -- de Vries equation and to the Kadomtsev -- Petviashvili equation. To find multiple soliton solutions for these well -- known equations author has used the Hirota method.

\end{abstract}

\section*{}

Recently Wazwaz \cite{WazwazCNSNS} studied the four M -- component nonlinear evolution equations, namely
the M -- component Korteweg -- de Vries (KdV) equation
\begin{equation}\begin{gathered}\label{Mur1}
\frac{\partial u_i}{\partial t}+\alpha_i\,\left(\sum_{k=1}^{M}\,u_k\right)\,\frac{\partial u_i}{\partial x}+\frac{\partial^3 u_i}{\partial x^3}=0,
\end{gathered}\end{equation}
the M -- component Kadomtsev -- Petviashvilli equation
\begin{equation}\begin{gathered}\label{Mur1a}
\left(\frac{\partial u_i}{\partial t}+\alpha_i\,\left(\sum_{k=1}^{M}\,u_k\right)\,\frac{\partial u_i}{\partial x}+\frac{\partial^3 u_i}{\partial x^3}\right)_x+\frac{\partial^2u_i}{\partial y^2}=0,
\end{gathered}\end{equation}
the M -- component mKdV equation
\begin{equation}\begin{gathered}\label{Mur1b}
\frac{\partial u_i}{\partial t}+\alpha_i\,\left(\sum_{k=1}^{M}\,u_{k}^{2}\right)\,\frac{\partial u_i}{\partial x}+\frac{\partial^3 u_i}{\partial x^3}=0,
\end{gathered}\end{equation}
and the M -- component mKdV -- KP equation
\begin{equation}\begin{gathered}\label{Mur1c}
\left(\frac{\partial u_i}{\partial t}+\alpha_i\,\left(\sum_{k=1}^{M}\,u_{k}^{2}\right)\,\frac{\partial u_i}{\partial x}+\frac{\partial^3 u_i}{\partial x^3}\right)_x+\frac{\partial^2u_i}{\partial y^2}=0.
\end{gathered}\end{equation}

Firstly author \cite{WazwazCNSNS} has aimed "to show that a variety of M -- component nonlinear evolution equations belong to the class of integrable equations". Secondly author \cite{WazwazCNSNS} has sought "to detrmine multiple soliton solutions and multiple -- singular soliton solutions for these equations".

The aim of this note is to show that author \cite{WazwazCNSNS} has not considered the M -- component nonlinear evolution equations. We demonstrate that using the additional condition for components $u_k$ author \cite{WazwazCNSNS} reduced the M -- component nonlinear evolution equations to the well -- known integrable equations.

Let us demonstrate this fact using the system of equations \eqref{Mur1}. Author \cite{WazwazCNSNS} looked for solutions of this system assuming
\begin{equation}\begin{gathered}\label{Mur2a}
u_i=R_i\,\left(\ln {f}\right)_{xx},
\end{gathered}\end{equation}

From condition \eqref{Mur2a} we obtain the following equalities
\begin{equation}\begin{gathered}\label{Mur3a}
\frac{u_1}{R_1}=\frac{u_2}{R_2}=\ldots\frac{u_k}{R_k}=\ldots\frac{u_M}{R_M}=\left(\ln {f}\right)_{xx},
\end{gathered}\end{equation}

Taking equations \eqref{Mur3a} into account we can present equation \eqref{Mur1} in the form
\begin{equation}\begin{gathered}\label{Mur4a}
\frac{\partial u_i}{\partial t}+\frac{\alpha_i}{R_i}\,\left(\sum_{k=1}^{M}\,R_k\right)\,u_i\,\frac{\partial u_i}{\partial x}+\frac{\partial^3 u_i}{\partial x^3}=0,
\end{gathered}\end{equation}

Assuming in \eqref{Mur3a} 
\begin{equation}\begin{gathered}\label{Mur5a}
u_i=\frac{6 R_i}{\alpha_i\,\sum_{k=1}^{M}R_k}\,u
\end{gathered}\end{equation}
we have the well -- known Korteweg -- de Vries equation in the form
\begin{equation}\begin{gathered}\label{Mur6a}
u_t+6\,u\,u_x+\,u_{xxx}=0
\end{gathered}\end{equation}

Note that equation \eqref{Mur5a} is the famous Korteweg - de Vries equation \cite{Korteweg, Kruskal, Ablowitz}. There are soliton solutions of this equation \cite{Gardner, Lax} that can be obtained by the Hirota method \cite{Hirota} taking the following formula into consideration
\begin{equation}\begin{gathered}\label{Mur8}
u=2\,\frac{\partial^2 \ln{f}}{\partial x^2}
\end{gathered}\end{equation}

As consequence of this observation we obtain that author \cite{WazwazCNSNS} did not study the system of equations \eqref{Mur1} but he looked for solution of the following system of equations
\begin{equation}\begin{gathered}\label{Mur9}
u_t+6\,u\,u_{x}+\,u_{xxx}=0, \quad u_i=\frac{6 R_i}{\alpha_i\,\sum_{k=1}^{M}R_k}\,u
\end{gathered}\end{equation}

The second equation is trivial algebraic equations for finding $u_i$. No doubt the system of equations \eqref{Mur9} is the completely integrable system but we do not see the subject of publication in this direction. We can suggest a lot of similar "completely integrable system of equations".

Let us note that using the formula \eqref{Mur5a} we can reduce \eqref{Mur1a} to the well - known Kadomtsev -- Petviashvili equation
\begin{equation}\begin{gathered}\label{Mur1aa}
\left(\frac{\partial u}{\partial t}+6\,u\,\frac{\partial u}{\partial x}+\frac{\partial^3 u}{\partial x^3}\right)_x+\frac{\partial^2 u}{\partial y^2}=0.
\end{gathered}\end{equation}
Soliton solutions of equation \eqref{Mur1aa} are well known \cite{Satsuma}.

Assuming in \eqref{Mur1b}
\begin{equation}\begin{gathered}\label{Mur5aa}
u_i=\frac{ R_{i}^2}{\alpha_i\,\sum_{k=1}^{M}R_{k}^2}\,v
\end{gathered}\end{equation}
we obtain the modified Korteweg -- de Vries equation
\begin{equation}\begin{gathered}\label{Mur1cc}
\frac{\partial v}{\partial t}+v^{2}\,\frac{\partial v}{\partial x}+\frac{\partial^3 v}{\partial x^3}=0.
\end{gathered}\end{equation}
Soliton solutions of \eqref{Mur1cc} are well known as well \cite{Ablowitz81}.

The the system of equation  \eqref{Mur1c} at $y=x$ can be reduced to the mKdV equation as well.

Unfortunately the author \cite{WazwazCNSNS} does not present new results to the problem of integrability of systems \eqref{Mur1}, \eqref{Mur1a}, \eqref{Mur1b} and \eqref{Mur1c} except trivial exercise on the application of the Hirota method to the famous nonlinear integrable equations. He has made the continuation of the errors that were discussed in recent papers \cite{Kudryashov08, Kudryashov09a, Kudryashov09b, Kudryashov09c, Kudryashov09d, Kudryashov09e, Kudryashov10a,  Kudryashov10b, Kudryashov10c, Parkes09b, Parkes09c}.

\end{document}